\documentstyle[bal,11pt]{article}
\title{
Computation of Higher-order Symmetries for Nonlinear
Evolution and Lattice Equations\thanks{Research supported in part 
by the NSF under Grant CCR-9625421.}
}
\author{%
\"{U}nal G\"{o}kta\c{s} 
and
Willy Hereman
\from
Department of Mathematical and Computer Sciences,\\
Colorado School of Mines, Golden, CO 80401-1887, U.S.A.\\
\email {{ugoktas,whereman}@mines.edu}.
}
\begin{document}
\maketitle
\begin{abstract}
A straightforward algorithm for the symbolic computation of higher-order 
symmetries of nonlinear evolution equations and lattice equations 
is presented.
The scaling properties of the evolution or lattice equations are used 
to determine the polynomial form of the higher-order symmetries.
The coefficients of the symmetry can be found by solving a linear system. 
The method applies to polynomial systems of PDEs of first-order in time
and arbitrary order in one space variable. Likewise, lattices must 
be of first order in time but may involve arbitrary shifts in the 
discretized space variable. 

The algorithm is implemented in {\it Mathematica} and can be used to
test the integrability of both nonlinear evolution equations 
and semi-discrete lattice equations. 
With our {\bf Integrability Package}, higher-order symmetries are obtained 
for several well-known systems of evolution and lattice equations. 
For PDEs and lattices with parameters, the code allows one to
determine the conditions on these parameters so that a sequence of
higher-order symmetries exist. The existence of a sequence of
such symmetries is a predictor for integrability. 

\amsmos
35Q53, 35R10, 34K99, 68Q40.


\end{abstract}

\section{Introduction}

A large number of physically important nonlinear models are completely
integrable, which means that they are linearizable via an explicit 
transformation or solvable via the Inverse Scattering Transform. 
Integrable continuous or discrete models arise in key branches of physics 
including classical, quantum, particle, statistical, and plasma physics. 
Integrable equations also model wave phenomena in nonlinear optics and 
the bio-sciences.
Mathematically, nonlinear models involve ordinary or partial differential 
equations (ODEs or PDEs), differential-difference equations (DDEs), 
integral equations, etc. \cite{AF1987}.

Whichever form they come in, completely integrable equations exhibit
analytic properties reflecting their rich mathematical structure. 
For instance, completely integrable PDEs and DDEs possess infinitely many 
symmetries and conserved quantities (if the model is conservative).
Perhaps after a suitable change of variables, the equations have the 
Painlev\'e property, admit B\"acklund transformations, prolongation 
structures, or can be written in bi-Hamiltonian form \cite{AF1987}.

The existence of an infinite hierarchy of symmetries for integrable 
equations can be established by explicitly constructing the recursion 
operator that connects such symmetries. 
Finding symmetries and recursion operators for nonlinear models is a 
nontrivial task, in particular if attempted with pen and paper. 
Computer algebra systems can greatly assist in the search for 
higher-order symmetries and recursion operators. 

In this paper we present a direct algorithm that allows one to 
automatically compute polynomial higher-order symmetries for polynomial 
PDEs in 1+1 dimension and polynomial DDEs (semi-discrete equations). 
The systems of DDEs or PDEs must be of evolution type, 
i.e. first order in (continuous) time. 
The number of equations and the order of differentiation 
(or shift level) in the spatial variable are arbitrary.

We use the dilation invariance of the given system of PDEs or DDEs 
to determine the form of the polynomial generalized symmetry. 
Upon substitution of the form of the symmetry into the defining equation, 
one has to solve a linear system for the unknown constant coefficients 
of the symmetry. In case the original system contains free parameters, 
the eliminant of that linear system will determine the necessary 
conditions for the parameters, so that the system admits the
required generalized symmetry. Our algorithm can thus be used 
as an integrability test for classes of equations involving parameters.

For the PDE case, a slight extension of our algorithm allows one to
compute higher-order symmetries that {\it explicitly} depend on the 
independent variables $x$ and $t$. 
However, in such cases, it is necessary to specify the highest 
degree of the independent variables in the generalized symmetry.

Once the generalized symmetries are explicitly known, it is quite often
possible to find the recursion operator by inspection \cite{AF1980}.
If the recursion operator is {\it hereditary} then the equation will
possess infinitely many symmetries. 
If the operator is hereditary and {\it factorizable} then the equation
has infinitely many conserved quantities \cite{AF1987,BFandWOandWW1987}. 

For Lagrangian systems the set of higher-order symmetries can be 
shown to lead to the set of conservation laws \cite{PO1977,VRandGK1994}. 
For equations without Lagrangian structure there is no universal 
correspondence between symmetries and conservation laws. 
For {\it Mathematica} algorithms and code to compute conservations laws 
of nonlinear PDEs and DDEs we refer to \cite{UGandWH1997a} 
and \cite{UGandWH1997b,UGandWHandGE1997}, respectively. 
The relationship between symmetries and conservation laws, as expressed 
through Noether's theorem, is beyond the scope of this paper 
(see \cite{VRandGK1994} for details). 

The paper is organized as follows. In Section 2 we give the definition of
generalized symmetry. We also give the three steps of the algorithm to 
compute generalized symmetries of nonlinear evolution equations. 
The Korteweg-de Vries and Boussinesq equations are used to illustrate
the computations. 
In Section 3 we extend the algorithm to nonlinear DDEs 
(the semi-discrete case), with the Toda lattice as leading example.  
The examples in Section 4 are meant to show how the technique
works for complex equations, like the nonlinear Schr\"odinger equation 
and one of its discretizations (the Ablowitz-Ladik system). 
We use a class of fifth-order KdV-type equations, and the 
Hirota-Satsuma system to exemplify the method for equations with parameters.
In Section 5 we give details about the use of our symmetry package 
in {\it Mathematica}. 
We briefly discuss other software for higher-order symmetries in Section 6. 
Conclusions and an outlook for future research are given in Section 7.

\section{Symmetries of Partial Differential Equations}

\subsection{Definition}

Consider a system of PDEs in the (single) space variable $x$ and time 
variable $t$,
\begin{equation}
\label{pdesys}
{\bf u}_t = {\bf F} ({\bf u}, {\bf u}_{x}, {\bf u}_{2x}, ..., {\bf u}_{mx}),
\end{equation}
where ${\bf u}$ and ${\bf F}$ are vector dynamical variables with
the same number of components: ${\bf u} = (u_1, u_2, ..., u_n), 
{\bf F} = (F_1, F_2, ..., F_n)$ and 
${\bf u}_{mx} = \frac{\partial^m {{\bf u}}}{\partial{x^m}}$.
The vector function ${\bf F}$ is assumed to be a polynomial function in  
${\bf u}, {\bf u}_x, ..., {\bf u}_{mx}$.  
There are no restrictions on the order of the system or its degree of
nonlinearity. 
If PDEs are of second or higher order in $t$, we assume that they
can be recast in the form (\ref{pdesys}).

A vector function ${\bf G} (x, t, {\bf u}, {\bf u}_{x}, {\bf u}_{2x}, ...),$
with ${\bf G} = (G_1, G_2, ..., G_n),$
is called a {\it symmetry\/} of (\ref{pdesys}) if and only if it leaves
(\ref{pdesys}) invariant for the replacement
${\bf u} \rightarrow {\bf u} + \epsilon {\bf G}$ 
within order $\epsilon.$ Hence, 
\begin{equation}
\label{invariance}
{\rm D}_t ({\bf u} + \epsilon {\bf G}) = 
{\bf F} ({\bf u} + \epsilon {\bf G})
\end{equation}
must hold up to order $\epsilon$ on the solutions of (\ref{pdesys}). 
Consequently, ${\bf G}$ must satisfy the linearized equation 
\cite{NEandWHS1992,AF1987}
\begin{equation}
\label{pdesymmetry}
{\rm D}_t {\bf G} = {\bf F}'({\bf u})[{\bf G}],
\end{equation}
where ${\bf F}'$ is the Fr\'echet derivative of {\bf F}, i.e.,
\begin{equation}
\label{pdefrechet}
{\bf F}'({\bf u})[{\bf G}] = 
{\partial \over \partial{\epsilon}} 
{\bf F}({\bf u}+\epsilon {\bf G})_{|{\epsilon = 0}} .
\end{equation}
\vskip 2pt
\noindent
In (\ref{invariance}) and (\ref{pdefrechet}) we infer that ${\bf u}$ 
is replaced by ${\bf u} + \epsilon {\bf G},$ and 
${\bf u}_{nx}$ by ${\bf u}_{nx} + \epsilon {\rm D}^n_x {\bf G}.$
As usual, ${\rm D}_{t}$ and ${\rm D}_{x}$ are total derivatives.

\noindent
Symmetries of the form ${\bf G} (x, t, {\bf u})$ are called 
{\it point\/} symmetries. 
If ${\bf G} (x, t, {\bf u}, {\bf u}_{x}, {\bf u}_{t})$ they are
called {\it classical \/} or Lie-B\"acklund symmetries, and all others 
symmetries involving higher derivatives than the first are called 
generalized or {\it higher-order\/} symmetries \cite{AMandASandVS1991}.

The examples used in the description of the algorithm, involve one or 
two dependent variables. For simplicity of notation, the components of 
${\bf u}$ will be denoted by $u,v, ...$ (instead of $u_1, u_2$, etc.). 

\subsection{Algorithm}

To illustrate our algorithm, we consider the Korteweg-de Vries (KdV) equation 
\begin{equation}
\label{kdv}
u_t = 6 u u_x + u_{3x}
\end{equation}
from soliton theory. This ubiquitous evolution equation is known to have
infinitely many symmetries \cite{AF1980}.

Key to our method is the observation that (\ref{kdv}) is invariant under
the dilation symmetry (or scaling)
\begin{equation}
\label{dilationkdv}
(t, x, u) \rightarrow ({\lambda}^{-3} t, \lambda^{-1} x, {\lambda}^{2} u),
\end{equation}
where $\lambda$ is an arbitrary parameter. The result of this dimensional 
analysis can be stated as follows: $u$ corresponds to two derivatives with 
respect to $x$, for short, $ u \sim \frac{{\partial}^2}{\partial{x^2}}$.
Similarly, 
$\frac{\partial}{\partial{t}} \sim \frac{{\partial}^3}{\partial{x^3}}$.
Scaling invariance, which is a special Lie-point symmetry, is an intrinsic
property of many integrable nonlinear PDEs and DDEs. 

Our algorithm exploits this scaling invariance to find symmetries, which now 
proceeds in three steps.
\vskip 5pt
\noindent
{\it Step 1: Determine the weights of variables}
\vskip 3pt
\noindent
The {\it weight\/}, $w$, of a variable is by definition equal to the number
of derivatives with respect to $x\/$ the variable carries.
Weights are rational, and weights of dependent variables are nonnegative.
We set $w(\frac{\partial}{\partial x}) = 1$.
In view of (\ref{dilationkdv}), we have $w(u) = 2$ and 
$w(\frac{\partial}{\partial t}) = 3. $
Consequently, $w(x) = -1$ and $w(t) = -3$.

The {\em rank} of a monomial is defined as the total weight of the monomial,
again in terms of derivatives with respect to $x$.
Observe that (\ref{kdv}) is an equation of rank 5, since all the terms 
(monomials) have the same rank, namely 5. 
This property is called {\it uniformity in rank}.

Conversely, requiring uniformity in rank for (\ref{kdv}) allows one 
to compute the weights of the dependent variables. 
Indeed, with $ w(\frac{\partial}{\partial x}) = 1$ we have
\[w(u) + w(\frac{\partial}{\partial t}) = 2 w(u) + 1 = w(u) + 3, \]
which yields
$ w(u) = 2, \, w(\frac{\partial}{\partial t}) = 3. $ 
Hence, $w(t) = -3,$ which is consistent with (\ref{dilationkdv}).
\vskip 5pt
\noindent
{\it Step 2: Construct the form of the symmetry}
\vskip 3pt
\noindent
As an example, let us compute the form of the symmetry of rank 7. 
Start by listing all powers in $u$ with rank $7$ or less:
$ {\cal L} \!=\! \{ 1, u, u^2, u^3 \}. $
Next, for each monomial in $\cal L$, introduce enough $x$-derivatives,
so that each term exactly has rank $7$. Thus, 
\[ 
{{\partial} \over \partial{x}} ( u^3 ) = 3 u^2 u_x, \quad
{{\partial}^3 \over \partial{x^3}} ( u^2 ) = 6 u_x u_{2x} + 2 u u_{3x}, \quad
{{\partial}^5 \over \partial{x^5}} ( u ) = u_{5x}, \quad
{{\partial}^7 \over \partial{x^7}} ( 1 ) = 0 .
\]
Then, gather the resulting (non-zero) terms in a set 
$
{\cal R} = \{u^2 u_x, u_x u_{2x}, u u_{3x}, u_{5x} \},
$
which contains the building blocks of the symmetry. 
Linear combination of the monomials in ${\cal R}$ with constant
coefficients $c_i$ gives the form of the symmetry:
\begin{equation}\label{formsym7kdv}
G = c_1 \, u^2 u_x + c_2 \, u_x u_{2x} + c_3 \, u u_{3x} + c_4 \, u_{5x} .
\end{equation}
\vskip 5pt
\noindent
{\it Step 3: Determine the unknown coefficients in the symmetry}
\vskip 3pt
\noindent
We determine the coefficients $c_i$ by requiring that 
(\ref{pdesymmetry}) holds on the solutions of (\ref{pdesys}).
Compute ${\rm D}_t {\bf G}$ and use (\ref{pdesys}) to remove ${\bf u}_t, 
{\bf u}_{tx}, {\bf u}_{txx},$ etc. 
For given ${\bf F},$ compute the Fr\'echet derivative (\ref{pdefrechet}) and, 
in view of (\ref{pdesymmetry}), equate the resulting expressions. 
Treating the different monomial terms in ${\bf u}$ and its $x$-derivatives 
as independent, the linear system for the coefficients $c_i$ is 
readily obtained. 

For (\ref{kdv}), we perform this computation with $F = 6 u u_x + u_{3x}$ 
and $G$ in (\ref{formsym7kdv}).  
Considering as independent all products and powers of $u, u_x, u_{xx},...,$ in 
\begin{eqnarray}
\label{sym7kdvresult}
\!\!\!\!\!\!\!\!\!\!\!&& 
(12 c_1 - 18 c_2) u_x^2 u_{2x} + (6 c_1 - 18 c_3) u u_{2x}^2 +
(6 c_1 - 18 c_3) u u_x u_{3x} + (3 c_2 - 60 c_4) u_{3x}^2 + \nonumber \\
\!\!\!\!\!\!\!\!\!\!\!&&
(3 c_2 + 3 c_3 - 90 c_4) u_{2x} u_{4x} + 
(3 c_3 - 30 c_4) u_x u_{5x} \equiv 0,
\end{eqnarray}
we obtain the linear system for the coefficients $c_i :$
\vskip 3pt
\noindent
$
{\cal S}=\{ 12 c_1 \!-\! 18 c_2 = 0, 6 c_1 \!-\! 18 c_3 = 0, 3 c_2 \!-\! 60 c_4 = 0, 
3 c_2 + 3 c_3 \!-\! 90 c_4 = 0, 3 c_3 \!-\! 30 c_4 = 0 \}.  
$
\vskip 3pt
\noindent
The solution is 
$ {c_1 \over 30} = {c_2 \over 20} = {c_3 \over 10} = c_4 $. 
Since symmetries can only be determined up to a multiplicative constant,
we choose $ c_1 = 30, c_2 = 20, c_3 = 10 $ 
and $ c_4 = 1 $, and substitute this into (\ref{formsym7kdv}).
Hence,
\[ 
G =  30 u^2 u_x + 20 u_x u_{2x} + 10 u u_{3x} + u_{5x}. 
\]   
Note that $u_t = G$ is known as the Lax equation, which is the fifth-order 
PDE in the completely integrable KdV hierarchy.
\vskip 3pt
\noindent
Analogously, for (\ref{kdv}) we computed the ($x\!-\!t$ independent) 
symmetries of rank $\leq 11.$
They are:
\begin{eqnarray*}
G^{(1)} \!&=&\! u_x, \;\;\;\; G^{(2)} = 6 u u_x + u_{3x}, \;\;\;\;
G^{(3)} = 30 u^2 u_x + 20 u_x u_{2x} + 10 u u_{3x} + u_{5x}, \\
G^{(4)} \!&=&\! 140 u^3 u_x + 70 u_x^3 + 280 u u_x u_{2x} + 70 u^2 u_{3x} 
+ 70 u_{2x} u_{3x} + 42 u_x u_{4x} \\
& & + 14 u u_{5x} + u_{7x}, \\
G^{(5)} &=& 630 u^4 u_x + 1260 u u_x^3 + 2520 u^2 u_x u_{2x} 
+ 1302 u_x u_{2x}^2 + 420 u^3 u_{3x} \\
& & + 966 u_x^2 u_{3x} 
+ 1260 u u_{2x} u_{3x} + 756 u u_x u_{4x} + 252 u_{3x} u_{4x} \\
& & + 126 u^2 u_{5x} 
+ 168 u_{2x} u_{5x} + 72 u_x u_{6x} + 18 u u_{7x} + u_{9x} .
\end{eqnarray*}
These results agree with those listed in the literature
(see e.g.\ \cite{AF1980,AMandASandVS1991,PO1993}).
\vskip 2pt
\noindent
{\bf Remarks.} 
\begin{itemize}
\item[(i)] 
The recursion operator \cite[p. 312]{PO1993} for the KdV equation is given by 
\begin{equation}
\label{recursion}
{\cal R} = D^2 + 4 u + 2 u_x D^{-1}.
\end{equation}
This operator is hereditary \cite{BFandWOandWW1987} and connects the 
above symmetries. 
\vskip 2pt
\noindent
For example,
\begin{eqnarray}
\label{recursionlinks}
{\cal R} u_x &=& (D^2 + 4 u + 2 u_x D^{-1}) u_x = 6 u u_x + u_{3x}, 
\nonumber \\
{\cal R} (6 u u_x + u_{3x}) &=& (D^2 + 4 u + 2 u_x D^{-1}) (6 u u_x + u_{3x}) 
\nonumber \\
&=&  30 u^2 u_x + 20 u_x u_{2x} + 10 u u_{3x} + u_{5x},   
\end{eqnarray}
and so forth. 
\item[(ii)] Instead of working with the definition 
(\ref{pdesymmetry})
of the symmetry, one could introduce an evolution equation, 
\begin{equation}
\label{pdeflowsymmetry}
{\bf u}_{\tau} = {\bf G} (x, t, {\bf u}, {\bf u}_{x}, {\bf u}_{2x}, ...),
\end{equation}
which defines the flow generated by ${\bf G}$ and parameterized by the 
auxiliary time variable $\tau$.
The symmetry can then be computed from the compatibility condition 
of (\ref{pdesys}) and (\ref{pdeflowsymmetry}): 
\begin{equation}
\label{pdecompatibility}
{\rm D}_\tau {\bf F} ({\bf u}, {\bf u}_{x}, {\bf u}_{2x}, ..., 
{\bf u}_{nx}) = 
{\rm D}_t {\bf G} (x, t, {\bf u}, {\bf u}_{x}, {\bf u}_{2x}, ...). 
\end{equation}
One then proceeds as follows:  
As above, determine the form of the symmetry ${\bf G}$ involving the constant 
coefficients $c_i.$ Then, compute ${\rm D}_t {\bf G}$ and use 
(\ref{pdesys}) to remove ${\bf u}_t, {\bf u}_{tx}$, etc.
Subsequently, compute ${\rm D}_\tau {\bf F} $ and use 
(\ref{pdeflowsymmetry}) to remove
${\bf u}_{\tau}, {\bf u}_{{\tau}x}$, etc. 
Finally, use (\ref{pdecompatibility}) to determine
the linear system for the unknown $c_i.$ Solve the system and
substitute the result into the form of ${\bf G}.$

Applied to our example, ${\rm D}_t G$ is computed with $G$ in 
(\ref{formsym7kdv}). 
Next, (\ref{kdv}) is used to eliminate all $t$-derivatives of $u$ from the 
expression of ${\rm D}_t G.$ 
Then, compute ${\rm D}_\tau F$ with $F$ in the right 
hand side of (\ref{kdv}), 
and eliminate all $\tau$-derivatives through (\ref{pdeflowsymmetry}) 
after substitution of (\ref{formsym7kdv}). 
Finally, expressing that ${\rm D}_\tau F - D_t G \equiv 0 $ leads to 
(\ref{sym7kdvresult}).

Although this procedure (see Ito \cite{MI1994}) circumvents the 
evaluation of the Fr\'echet derivative, it seems more involved 
than our algorithm which uses the definition (\ref{pdesymmetry}). 

\end{itemize}

\subsection{Symmetries Explicitly Dependent on $x$ and $t$}

The KdV equation (\ref{kdv}) has also symmetries which {\it explicitly} 
depend on $x$ and $t$. Our algorithm can be used to find these symmetries
provided that we specify the maximum degree in $x$ and $t.$ 

As an example, we will compute the symmetry of rank 2 for (\ref{kdv}), 
that linearly depends on $x$ and/or $t.$ In other words, the highest 
degree in $x$ or $t$ in the symmetry is $1$. 

We start with the list of monomials in $u, x$ and $t$ of rank $2$ or less:
\[ 
{\cal L} = \{ 1, u, x, x u, t, t u, t u^2 \}. 
\]
Then, for each monomial in $\cal L$, introduce enough $x$-derivatives,
so that each term exactly has weight $2$. Thus,
\[ 
{\rm D}_x ( x u ) = u + x u_x, \;
{\rm D}_x ( t u^2 ) = 2 t u u_x, \;
{\rm D}^3_x ( t u ) = t u_{3x}, \;
{\rm D}^2_x ( 1 ) = {\rm D}^3_x ( x ) = {\rm D}^5_x ( t ) = 0 .
\]
Gather the non-zero resulting terms in a set
$
{\cal R} = \{u, x u_x, t u u_x, t u_{3x} \},
$
which contains the building blocks of the symmetry.
Linear combination of the monomials in ${\cal R}$ with constant
coefficients $c_i$ gives the form of the symmetry:
\begin{equation}\label{formsym2kdv}
G = c_1 \, u + c_2 \, x u_x + c_3 \, t u u_x + c_4 \, t u_{3x} .
\end{equation}
Now, determine the coefficients $c_1$ through $c_4$ by requiring that
(\ref{pdesymmetry}) holds on the solutions of (\ref{kdv}). 
After grouping the terms, one gets
\[
(6 c_1 + 6 c_2 - c_3) u u_x + (3 c_3 - 18 c_4) t u_{2x}^2 +
(3 c_2 - c_4) u_{3x} + (3 c_3 - 18 c_4) t u_x u_{3x}
\equiv 0,
\]
which yields
\[
{\cal S} = \{ 6 c_1 + 6 c_2 - c_3 = 0, 
3 c_3 - 18 c_4 = 0, 3 c_2 - c_4 = 0 \}.
\]
The solution is
$ {3 c_1 \over 2} = { 3 c_2 } = {c_3 \over 6} = c_4 $.
We choose $ c_1 = {2 \over 3}, c_2 = {1 \over 3}, c_3 = 6 $
and $ c_4 = 1 $, and substitute this into (\ref{formsym2kdv}).
Hence,
\[
G =  {2 \over 3} u + {1 \over 3} x u_x + 6 t u u_x + t u_{3x} .
\]
Similarly, we computed other symmetries of (\ref{kdv}) that linearly
depend on $x$ and $t$. They are of rank $0$ and $2:$ 
\[
G = 1 + 6 t u_x, \;\;\;{\rm and}\;\;\;
G = 2 u + x u_x + t u_t = 2 u + x u_x + 6 t u u_x + t u_{3x}.
\]
Our results agree with those in the literature \cite{AMandASandVS1991}.

\subsection{Example: Boussinesq equation}

For scaling invariant systems such as (\ref{kdv}), it suffices to consider
the dilation symmetry on the space of independent and dependent variables.
For systems that are inhomogeneous under a suitable scaling symmetry, 
such as the example given below, we use the following trick:
We introduce one (or more) auxiliary parameter(s) with an appropriate 
scaling. These extra parameters can be viewed as additional 
dependent variables, however, their derivatives are zero. 
By extending the action of the dilation symmetry to the space of 
independent and dependent variables, {\it including} the parameters, we are 
able to apply our algorithm to a larger range of polynomial PDE systems.

Consider the wave equation,
\begin{equation} \label{Bouseq}
u_{tt} - u_{2x} + 3 u u_{2x} + 3 u_x^2 + \alpha u_{4x} = 0,
\end{equation}
($\alpha$ constant) which was proposed by Boussinesq to describe 
surface water waves whose horizontal scale is much larger than the 
depth of the water \cite{MAandPC1991}.

To apply our algorithm, we must first rewrite (\ref{Bouseq}) as a 
first-order system,
\begin{eqnarray} 
\label{Boussys1}
u_t &=& v_x , \nonumber \\
v_t &=& u_x - 3 u u_x - \alpha u_{3x},
\end{eqnarray}
where $v$ is an auxiliary dependent variable.
It is easy to verify that the terms $u_x$ and $\alpha u_{3x}$ in the second
equation obstruct uniformity in rank.
To circumvent the problem we introduce an auxiliary parameter $\beta$ with
(unknown) weight, and replace (\ref{Boussys1}) by
\begin{eqnarray} 
\label{Boussys2}
u_t &=& v_x ,\nonumber \\
v_t &=& \beta u_x - 3 u u_x - \alpha u_{3x} .
\end{eqnarray}
As described in Step 1 we compute the weights from
\begin{eqnarray*}
& & w(u)+w(\frac{\partial}{\partial t}) = w(v)+1, \\
& & w(v)+w(\frac{\partial}{\partial t}) 
= w(\beta) + w(u)+1 = 2 w(u)+1 = w(u)+3.
\end{eqnarray*}
This yields
\[ w(u) = 2, \; w(v) = 3, \; w(\beta) = 2, \;\; {\rm and} \;
w(\frac{\partial}{\partial t}) = - w(t) = 2, \]
and the scaling properties of (\ref{Boussys2}) are
$ u \sim \beta \sim \frac{\partial{}}{\partial{t}} \sim 
 {\partial^2 \over \partial x^2}, \, v \sim {\partial^3 \over \partial x^3}.
$
Indeed, (\ref{Boussys2}) is invariant under the dilation symmetry
\[
(x, t, u, v, \beta)
\rightarrow
(\lambda^{-1} x, \lambda^{-2} t, \lambda^{2} u,
\lambda^{3} v, \lambda^{2} \beta).
\]
Observe that all the monomials in the equations in (\ref{Boussys2}) have 
rank $4$ and $5$.
Therefore, for any symmetry ${\bf G} $ of (\ref{Boussys2}), 
\[
{\rm rank}(G_2) = {\rm rank}(G_1) + 1 = {\rm rank}(G_1) + w(v)-w(u).
\]
Let us construct the form of the symmetry ${\bf G} = (G_1, G_2)$ with
${\rm rank}(G_1)=6$ and ${\rm rank}(G_2)=7.$
First, list all monomials in $u, v$ and $\beta$ of rank $6$ (respectively 
rank $7$) or less:
\begin{eqnarray*}
{\cal L}_1 &=& \{ 1, \beta, {\beta}^2, {\beta}^3, u, \beta u, {\beta}^2 u,
                  u^2, \beta u^2, u^3, v, \beta v, u v, v^2 \}, \\
{\cal L}_2 &=& \{ 1, \beta, {\beta}^2, {\beta}^3, u, \beta u, {\beta}^2 u,
                  u^2, \beta u^2, u^3, v, \beta v, {\beta}^2 v, u v,
                  \beta u v, u^2 v, v^2 \}.
\end{eqnarray*}
Next, for each monomial in ${\cal L}_1$ and ${\cal L}_2$, 
introduce the necessary $x$-derivatives,
so that each term in ${\cal L}_1$ exactly has rank $6$, and
each term in ${\cal L}_2$ has rank $7$.
Keeping in mind that $\beta$ is constant, and proceeding with the 
rest of the algorithm, we obtain:
\begin{eqnarray}
\label{boussys2sym67}
G_1^{(1)} &\!\!=\!\!& u_x v + u v_x + \frac{2}{3} \alpha v_{3x}, 
\nonumber \\
G_2^{(1)} &\!\!=\!\!& \beta u u_x - 3 u^2 u_x + v v_x - 6 \alpha u_x u_{2x} +
\frac{2}{3} \alpha \beta u_{3x} - 3 \alpha u u_{3x} - 
\frac{2}{3} {\alpha}^2 u_{5x}. 
\end{eqnarray}
Finally, setting $\beta = 1$ in (\ref{boussys2sym67}), one obtains a
symmetry of (\ref{Boussys1}) although initially this system was not
uniform in rank. We list one more higher-order symmetry of (\ref{Boussys1}):
\begin{eqnarray*}
G_1^{(2)} &=& u u_x - \frac{3}{2} u^2 u_x + v v_x - 5 \alpha u_x u_{2x}
+ \frac{2}{3} \alpha u_{3x} - 2 \alpha u u_{3x} 
- \frac{8}{15} {\alpha}^2 u_{5x}, \\
G_2^{(2)} &=& u v_x + v u_x - 3 u u_x v - \frac{3}{2} u^2 v_x
- 2 \alpha u_{2x} v_x - 3 \alpha u_x v_{2x} - \alpha u_{3x} v
+ \frac{2}{3} \alpha v_{3x} \\
& & - 2 \alpha u v_{3x} - \frac{8}{15} {\alpha}^2 v_{5x}.
\end{eqnarray*} 

\section{Symmetries of Differential-difference Equations}

\subsection{Definition}

Consider a system of DDEs, 
\begin{equation}
\label{ddesys}
{\dot{\bf u}}_n =
{\bf F} (...,{\bf u}_{n-1}, {\bf u}_{n}, {\bf u}_{n+1},...),
\end{equation}
where the equations are continuous in time, and discretized in the 
(single) space variable.
As before, ${\bf u}_{n}$ and ${\bf F}$ are vector dynamical variables 
with any number of components, and ${\bf F}$ is assumed to be a
polynomial with constant coefficients. 
There are no restrictions on the level of shifts or the degree of 
nonlinearity.
If DDEs are of second or higher order in $t$, they must be recast in 
the form (\ref{ddesys}).

A vector function
${\bf G} (...,{\bf u}_{n-1}, {\bf u}_{n}, {\bf u}_{n+1},...) $
is called a {\it symmetry\/} of (\ref{pdesys}) if the infinitesimal
transformation
\begin{equation}
\label{ddetransf}
{\bf u}_n \rightarrow {\bf u}_n +
\epsilon {\bf G} (...,{\bf u}_{n-1}, {\bf u}_{n}, {\bf u}_{n+1},...) 
\end{equation}
leaves (\ref{ddesys}) invariant within order $\epsilon$. 

Consequently, ${\bf G}$ must satisfy the linearized equation 
\cite{NEandWHS1992,AF1987}
\begin{equation}
\label{ddesymmetry}
{\rm D}_t {\bf G} = {\bf F}'({\bf u}_n)[{\bf G}],
\end{equation}
where ${\bf F}'$ is the Fr\'echet derivative of {\bf F}, defined as
\begin{equation}
\label{ddefrechet}
{\bf F}'({\bf u}_n)[{\bf G}] = 
{\partial \over \partial{\epsilon}} 
{\bf F}({\bf u}_n +\epsilon {\bf G})_{|{\epsilon = 0}} .
\end{equation}
\vskip 2pt
\noindent
Of course, (\ref{ddetransf}) means that 
${\bf u}_{n+k}$ is replaced by 
${\bf u}_{n+k} + \epsilon {{\bf G}_{|{n \rightarrow n+k}}}$.
For compactness of notation, in (\ref{ddesymmetry}) and 
(\ref{ddefrechet}) we used ${\bf F}'({\bf u}_n)$ instead of
${\bf F}'(...,{\bf u}_{n-1}, {\bf u}_n, {\bf u}_{n+1}, ... ).$

Also for notational simplicity, in the description of the algorithm below,
the components of ${\bf u}_n$ will be denoted by $u_n, v_n$, etc. 
We use $F_1, F_2, ...$ and $G_1, G_2, ...$,
to denote the components of $\bf F$ and $\bf G,$ respectively.

\subsection{Algorithm}

As the leading example, we consider the one-dimensional lattice
\cite{MH1974,MT1981}
\begin{equation}\label{orgtoda}
{\ddot{y}}_n = \exp{(y_{n-1} - y_n)} - \exp{(y_n - y_{n+1})},
\end{equation}
due to Toda.
In (\ref{orgtoda}), $y_n$ is the displacement from equilibrium of the
$n\/$th particle with unit mass under an exponential decaying interaction
force between nearest neighbors.
With the change of variables,
\[
u_n = {\dot{y}}_n, \quad\quad\quad  v_n = \exp{(y_{n} - y_{n+1})},
\]
the Toda lattice (\ref{orgtoda}) can be written in polynomial form
\begin{equation}
\label{todalatt}
{\dot{u}}_n = v_{n-1} - v_n, \quad\;\;\; {\dot{v}}_n = v_n (u_n - u_{n+1}).
\end{equation}
Observe that (\ref{todalatt}) is invariant under the dilation symmetry
\begin{equation}\label{dilationtoda}
(t, u_n, v_n) \rightarrow (\lambda^{-1} t, \lambda u_n, {\lambda}^{2} v_n),
\end{equation}
where $\lambda$ is an arbitrary parameter. 
Thus, $u_n$ corresponds to one derivative with respect to $t,$ or
$ u_n \sim \frac{\rm{d}}{\rm{dt}},$ 
and, similarly, $v_n \sim \frac{{\rm{d}}^2}{{\rm{dt}}^2} $.
Our 3-step algorithm exploits this scaling property to find symmetries. 
\vskip 5pt
\noindent
{\em Step 1: Determine the weights of variables}
\vskip 2pt
\noindent
In contrast to the algorithm for PDEs, we have to define
the {\it weight\/}, $w$, of variables in terms of the number
of derivatives with respect to $t,$ and we set
$w(\frac{\rm{d}}{\rm{dt}}) = 1$. 
Weights of dependent variables are nonnegative, rational, and independent 
of $n$.
In view of (\ref{dilationtoda}), we have $w(u_n) = 1$, and $w(v_n) = 2$.

The {\it rank} of a monomial is defined as the total weight of the monomial,
again in terms of derivatives with respect to $t$.
Observe that in the first equation of (\ref{todalatt}), all the
monomials have the same rank, namely 2, and in the second equation,
all the terms have rank 3. 

Conversely, requiring uniformity in rank for each equation in
(\ref{todalatt}) allows one to compute the weights of the dependent
variables. Indeed,
\[
w(u_n) +1 \!=\! w(v_n), \quad w(v_n) +1 \!=\! w(u_n) + w(v_n), 
\]
yields
$ w(u_n) = 1, \, w(v_n) = 2, $ which is consistent with (\ref{dilationtoda}).
\vskip 5pt
\noindent
{\em Step 2: Construct the form of the symmetry}
\vskip 2pt
\noindent
As an example, we compute the form of the symmetry of rank $(3,4)$,
i.e.\ $G_1$ and $G_2$ will have ranks 3 and 4, respectively.
Start by listing all monomials in $u_n$ and $v_n$ of ranks 
$3$ and 4, or less: 
\begin{eqnarray}
{\cal L}_1 &=& \{ u_n^3, u_n^2, u_n v_n, u_n, v_n \}, \nonumber \\
{\cal L}_2 &=& 
\{ u_n^4, u_n^3, u_n^2 v_n, u_n^2, u_n v_n, u_n, v_n^2, v_n \}.
\nonumber 
\end{eqnarray}
Next, for each monomial in ${\cal L}_1$ and ${\cal L}_2$, 
introduce the necessary $t$-derivatives.
so that each term exactly has rank $3$ and $4$, respectively. 
At the same time, use (\ref{todalatt}) to remove all $t-$derivatives.
Doing so, based on ${\cal L}_1,$ we obtain
\begin{eqnarray*}
&&
\!\!\!\!\!
{{\rm{d}}^0 \over {\rm{dt}}^0} ( u_n^3 )
   = u_n^3 , \;\;\;\;\quad
{{\rm{d}}^0 \over {\rm{dt}}^0} ( u_n v_n )
   = u_n v_n , \\
&& \!\!\!\!\!  \\
&& \!\!\!\!\!
{{\rm{d}} \over {\rm{dt}}} ( u_n^2 )
   = 2 u_n {\dot{u}}_n = 2 u_n v_{n-1} - 2 u_n v_n ,  \;\;\;\;\quad
{{\rm{d}} \over {\rm{dt}}} ( v_n )
   = {\dot{v}}_n =  u_n v_n -  u_{n+1} v_n, \\
&&  \!\!\!\!\!  \\
&&
\!\!\!\!\!
{{\rm{d}}^2 \over {\rm{dt}}^2} ( u_n )
   = {{\rm{d}} \over {\rm{dt}}} ( {\dot{u}}_n )
   = {{\rm{d}} \over {\rm{dt}}} ( v_{n-1} - v_n )
   = u_{n-1} v_{n-1} - u_{n} v_{n-1} - u_n v_n + u_{n+1} v_n .
\end{eqnarray*}
Gather the resulting terms in a set: 
$
{\cal R}_1 = 
\{ u_n^3, u_{n-1} v_{n-1} , u_n v_{n-1} , u_n v_n , u_{n+1} v_n \} .
$
\vskip 2pt
\noindent
Similarly, based on the monomials in ${\cal L}_2,$ we get
\begin{eqnarray*}
{\cal R}_2 & = &
\{ u_n^4, u_{n-1}^2 v_{n-1}, u_{n-1} u_n v_{n-1}, u_n^2 v_{n-1},
v_{n-2} v_{n-1}, v_{n-1}^2, u_n^2 v_n, \\
& &  u_n u_{n+1} v_n, u_{n+1}^2 v_n,
v_{n-1} v_n, v_n^2, v_n v_{n+1} \} .
\end{eqnarray*}
Linear combination of the monomials in ${\cal R}_1$ and ${\cal R}_2$ 
with constant coefficients $c_i$ gives the explicit form of the symmetry:
\begin{eqnarray}
\label{formsym3toda}
G_1 &=& c_1 \, u_n^3 + c_2 \, u_{n-1} v_{n-1} + c_3 \, u_n v_{n-1} +
c_4 \, u_n v_n + c_5 \, u_{n+1} v_n, \nonumber \\
G_2 &=& c_6 \, u_n^4 + c_7 \, u_{n-1}^2 v_{n-1} + 
c_8 \, u_{n-1} u_n v_{n-1} + c_9 \, u_n^2 v_{n-1} +
c_{10} \, v_{n-2} v_{n-1} + \nonumber \\ 
& & c_{11} \, v_{n-1}^2 + c_{12} \, u_n^2 v_n +
c_{13} \, u_n u_{n+1} v_n + c_{14} \, u_{n+1}^2 v_n +
c_{15} \, v_{n-1} v_n + \nonumber \\
& & c_{16} \, v_n^2 + c_{17} \, v_n v_{n+1} .
\end{eqnarray} 
\vskip 5pt
\noindent
{\em Step 3: Determine the unknown coefficients in the symmetry}
\vskip 2pt
To determine the coefficients $c_i$ we require that (\ref{ddesymmetry}) 
holds on any solution of (\ref{ddesys}).
Compute ${\rm D}_t {\bf G}$ and use (\ref{ddesys}) to remove all 
$ {\dot {\bf u}}_{n-1}, {\dot {\bf u}}_n,
{\dot {\bf u}}_{n+1},$ etc. 
Compute the Fr\'echet derivative (\ref{ddefrechet}) and, in view of 
(\ref{ddesymmetry}), equate the resulting expressions. 
Considering as independent all the monomials in 
${\bf u}_n $ and their shifts, we obtain the linear system that
determines the coefficients $c_i.$

Applied to (\ref{todalatt}) with (\ref{formsym3toda}), we obtain the solution
\begin{eqnarray}
\label{sym3todaresult}
c_1 = c_6 = c_7 = c_8 = c_9 = c_{10} = c_{11} = c_{13} = c_{16} = 0, \\
-c_2 = -c_3 = c_4 = c_5 = -c_{12} = c_{14} = -c_{15} = c_{17} .
\end{eqnarray}
Therefore, with the choice $c_{17} = 1,$ the symmetry is
\begin{eqnarray}\label{todasym34}
G_1 &=& u_n v_n - u_{n-1} v_{n-1} + u_{n+1} v_n - u_{n} v_{n-1}, \nonumber \\
G_2 &=& u_{n+1}^2 v_n - u_n^2 v_n + v_n v_{n+1} - v_{n-1} v_n.  
\end{eqnarray}
It is easy to produce new completely integrable DDEs based on these
symmetries. For instance, the DDE system
\begin{eqnarray}
\label{newintsys}
{\dot{u}}_n &=& G_1 = 
u_n v_n - u_{n-1} v_{n-1} + u_{n+1} v_n - u_{n} v_{n-1}, \nonumber \\
{\dot{v}}_n &=& G_2 = u_{n+1}^2 v_n - u_n^2 v_n + v_n v_{n+1} - v_{n-1} v_n.  
\end{eqnarray}
is also completely integrable. 

To illustrate the effectiveness of our algorithm to filter out
integrable cases among systems of DDEs with parameters, consider
a parameterized version of the Toda lattice, 
\begin{equation}\label{partodalatt}
{\dot{u}}_n = \alpha \; v_{n-1} - v_n, \quad
{\dot{v}}_n = v_n \; (\beta \; u_n - u_{n+1}),
\end{equation}
where $\alpha$ and $\beta$ are {\it nonzero} constants.
In \cite{ARandBGandKT1992} it was shown that (\ref{partodalatt})
is completely integrable if and only if $\alpha = \beta = 1.$

Using our algorithm, one can easily compute the {\it compatibility conditions}
for $\alpha$ and $\beta$, so that (\ref{partodalatt}) admits a polynomial
symmetry, say, of rank $(3,4)$.
The steps are the same as for (\ref{todalatt}).
However, the linear system for the $c_i$ is parameterized by $\alpha$ and 
$\beta$ and must be analyzed carefully. 
This analysis leads to the condition $ \alpha = \beta = 1.$ 

For $ \alpha = \beta = 1, $ (\ref{partodalatt}) coincides with 
(\ref{todalatt}), for which we computed symmetries with ranks $(4,5)$ 
and $(5,6).$ They are:
\begin{eqnarray*}
G_1^{(1)} &=& u_n^2 v_n + u_n u_{n+1} v_n + u_{n+1}^2 v_n + v_n^2 
+ v_n v_{n+1} - u_{n-1}^2 v_{n-1} - u_{n-1} u_n v_{n-1} \\
& & - u_n^2 v_{n-1} - v_{n-2} v_{n-1} - v_{n-1}^2, \\
G_2^{(1)} &=& u_{n+1} v_n^2 + 2 u_{n+1} v_n v_{n+1} + u_{n+2} v_n v_{n+1}
- u_n^3 v_n + u_{n+1}^3 v_n \\
& & - u_{n-1} v_{n-1} v_n - 2 u_n v_{n-1} v_n - u_n v_n^2, \\
G_1^{(2)} &=& u_n^3 v_n + u_n^2 u_{n+1} v_n + u_n u_{n+1}^2 v_n
+ u_{n+1}^3 v_n + 2 u_n v_n^2 + 2 u_{n+1} v_n^2 \\
& & + u_n v_n v_{n+1}
+ 2 u_{n+1} v_n v_{n+1} + u_{n+2} v_n v_{n+1} - u_{n-1}^3 v_{n-1}
- u_{n-1}^2 u_n v_{n-1} \\
& & - u_{n-1} u_n^2 v_{n-1} - u_n^3 v_{n-1}
- u_{n-2} v_{n-2} v_{n-1} - 2 u_{n-1} v_{n-2} v_{n-1} \\
& & - u_n v_{n-2} v_{n-1} - 2 u_{n-1} v_{n-1}^2 - 2 u_n v_{n-1}^2, \\
G_2^{(2)} &=& u_{n+1}^4 v_n - u_n^4 v_n - u_{n-1}^2 v_{n-1} v_n
- 2 u_{n-1} u_n v_{n-1} v_n - 3 u_n^2 v_{n-1} v_n \\ 
& & - v_{n-2} v_{n-1} v_n 
- v_{n-1}^2 v_n - 2 u_n^2 v_n^2 + 2 u_{n+1}^2 v_n^2 - v_{n-1} v_n^2
+ 3 u_{n+1}^2 v_n v_{n+1} \\
& & + 2 u_{n+1} u_{n+2} v_n v_{n+1}
+ u_{n+2}^2 v_n v_{n+1} + v_n^2 v_{n+1} + v_n v_{n+1}^2 + v_n v_{n+1} v_{n+2}.
\end{eqnarray*}

\section{More Examples}

\subsection{Nonlinear Schr\"{o}dinger Equation}

The nonlinear Schr\"{o}dinger (NLS) equation \cite{MAandPC1991},
\begin{equation}\label{NLS}
i q_t - q_{2x} + 2 {|q|}^2 q =0,
\end{equation}
arises as an asymptotic limit of a slowly varying dispersive wave envelope
in a nonlinear medium, and as such has significant applications in
nonlinear optics, water waves, and plasma physics.
Together with the ubiquitous KdV equation (\ref{kdv}), the completely 
integrable NLS equation is one of the most studied soliton equations.

In order to compute the symmetries of (\ref{NLS}) we consider $q$ 
and $q^{*}$ as independent
variables and add the complex conjugate equation to (\ref{NLS}).
Absorbing $i$ in the scale of $t,$ we get
\begin{equation}\label{NLSsys}
q_t - q_{2x} + 2 q^2 q^{*} = 0, \quad
q_t^{*} + q_{2x}^{*} - 2 q^{*2} q = 0.
\end{equation}
Since $w(q) = w(q^{*})$, we obtain 
\[
w(q) = w(q^{*}) = 1, \;\;\;{\rm and}\;\;\;
w(\frac{\partial}{\partial t}) = - w(t) = 2.
\]
Hence, the following are the symmetries of ranks $(4,4), (5,5)$, and 
$(6,6)$: 
\begin{eqnarray}
\label{nlssymmetries}
G_1^{(1)} &=& - 6 q q_x q^{*} + q_{3x}, 
\quad\;\; G_2^{(1)} = {G^*_1}^{(1)} = -6 q q^{*} q^{*}_x + q^{*}_{3x}, \nonumber \\
G_1^{(2)} &=& -6 q^3 {q^{*}}^2 + 6 q_x^2 q^{*}
+ 4 q q_x q^{*}_x + 8 q q_{2x} q^{*} 
+ 2 q^2 q^{*}_{2x} - q_{4x}, \quad G_2^{(2)} = {G^*_1}^{(2)} \nonumber \\
G_1^{(3)} &=& 30 q^2 q_x {q^{*}}^2 - 10 {q_x}^2 q^{*}_x 
- 20 q_x q_{2x} q^{*} 
- 10 q q_{2x} q^{*}_x - 10 q q_x q^{*}_{2x} \nonumber \\
& & - 10 q q_{3x} q^{*} + q_{5x}, 
\quad\;\; G_2^{(3)} = {G^*_1}^{(3)}. \nonumber 
\end{eqnarray}

\subsection{Fifth-Order Korteweg-de Vries Equations}

Consider the parameterized family of fifth-order equations,
\begin{equation} 
\label{kdv5par}
u_t + \alpha u^2 u_{x} + \beta u_x u_{2x} + \gamma u u_{3x} + u_{5x} = 0,
\end{equation}
where $\alpha,\beta,\gamma$ are nonzero constants.
Integrable cases of (\ref{kdv5par}) are well known in the literature
\cite{AFandJG1980,RHandMI1983,BKandGW1981,JSandDK1977}. 
Indeed, for $\alpha = 30, \beta = 20, \gamma = 10,$ equation (\ref{kdv5par})
reduces to the Lax equation \cite{PL1968}.
The SK equation, due to Sawada and Kotera \cite{KSandTK1974}, and 
Dodd and Gibbon \cite{RDandJG1977}, is obtained for 
$\alpha = 5, \beta = 5, \gamma = 5.$
The KK equation, due to Kaup \cite{DK1980} and Kupershmidt, corresponds to
$\alpha = 20, \beta = 25, \gamma = 10$. 

The scaling properties of (\ref{kdv5par}) are such that
$ u \sim \frac{\partial^2}{\partial{x^2}},\quad
\frac{\partial}{\partial{t}} \sim \frac{\partial^5}{\partial{x^5}}. 
$
Using our algorithm, one easily computes the {\it compatibility conditions}
for the parameters $\alpha, \beta$ and $\gamma,$ so that (\ref{kdv5par})
admits a symmetry of fixed rank. The results are: 
\vskip 5pt
\noindent
{\bf Rank 3:} $ G = u_x$ is a symmetry of (\ref{kdv5par}) without any
conditions on the parameters. 
\vskip 5pt
\noindent
{\bf Rank 5:} $ G = u u_x + \frac{5}{3 \gamma} u_{3x}$ is a symmetry of
(\ref{kdv5par}) provided that
\begin{equation}\label{laxcond}
\alpha = \frac{3}{10} {\gamma}^2, \;{\rm and}\; \beta = 2 \gamma .
\end{equation}
The Lax equation satisfies (\ref{laxcond}).
Since the KdV equation (\ref{kdv}) is a member of Lax hierarchy, condition
(\ref{laxcond}) comes as no surprise. 
\vskip 5pt
\noindent
{\bf Rank 7:} Equation (\ref{kdv5par}) is of rank $7$. 
The stationary part of (\ref{kdv5par}) is the symmetry. 
\vskip 5pt
\noindent
{\bf Rank 9:} Three branches emerge:
\begin{enumerate}
\item[(i)] If condition (\ref{laxcond}) holds then
\begin{eqnarray*} 
G &\!=\!&\! u^3 u_x + \frac{5}{\gamma} u_x^3 + \frac{20}{\gamma} u u_x u_{2x} 
+ \frac{5}{\gamma} u^2 u_{3x} + \frac{50}{{\gamma}^2} u_{2x} u_{3x}
+ \frac{30}{{\gamma}^2} u_x u_{4x} \\
&&\! + \frac{10}{{\gamma}^2} u u_{5x} + \frac{50}{7 {\gamma}^3} u_{7x} .
\end{eqnarray*}
\item[(ii)] If 
\begin{equation}\label{skcond}
\alpha = \frac{1}{5} {\gamma}^2, \;{\rm and}\; \beta = \gamma
\end{equation}
holds, one has the symmetry
\begin{eqnarray*}
G &\!=\!&\! u^3 u_x + \frac{15}{4 \gamma} u_x^3 
+ \frac{45}{2 \gamma} u u_x u_{2x} + \frac{15}{2 \gamma} u^2 u_{3x} 
+ \frac{225}{4 {\gamma}^2} u_{2x} u_{3x}
+ \frac{75}{2 {\gamma}^2} u_x u_{4x} \\
&&\! + \frac{75}{4 {\gamma}^2} u u_{5x}
+ \frac{375}{28 {\gamma}^3} u_{7x} .
\end{eqnarray*}
The SK equation satisfies the condition (\ref{skcond}).
\item[(iii)] One has the symmetry
\begin{eqnarray*}
G &=& u^3 u_x + \frac{75}{8 \gamma} u_x^3 + \frac{135}{4 \gamma} u u_x u_{2x}
+ \frac{15}{2 \gamma} u^2 u_{3x} + \frac{225}{2 {\gamma}^2} u_{2x} u_{3x}
+ \frac{525}{8 {\gamma}^2} u_x u_{4x} \\
& & + \frac{75}{4 {\gamma}^2} u u_{5x}
+ \frac{375}{28 {\gamma}^3} u_{7x} 
\end{eqnarray*}
provided that
\begin{equation}\label{kkcond}
\alpha = \frac{1}{5} {\gamma}^2, \;{\rm and}\; \beta = \frac{5}{2} \gamma , 
\end{equation}
which holds for the KK case.
\end{enumerate}
{\bf Rank 11:} One obtains the symmetry 
\begin{eqnarray*}
G &=& u^4 u_x + \frac{20}{\gamma} u u_x^3 + \frac{40}{\gamma} u^2 u_x u_{2x}
+ \frac{620}{3 {\gamma}^2} u_x u_{2x}^2 + \frac{20}{3 \gamma} u^3 u_{3x}
+ \frac{460}{3 {\gamma}^2} u_x^2 u_{3x} \\
& & + \frac{200}{{\gamma}^2} u u_{2x} u_{3x}
+ \frac{120}{{\gamma}^2} u u_{x} u_{4x}
+ \frac{400}{{\gamma}^3} u_{3x} u_{4x}
+ \frac{20}{{\gamma}^2} u^2 u_{5x} 
+ \frac{800}{3 {\gamma}^3} u_{2x} u_{5x} \\
& & + \frac{800}{7 {\gamma}^3} u_x u_{6x}
+ \frac{200}{7 {\gamma}^3} u u_{7x} + \frac{1000}{63 {\gamma}^4} u_{9x}
\end{eqnarray*}
provided that the condition (\ref{laxcond}) for the Lax hierarchy 
is satisfied.

In summary, our algorithm allows one to filter out all the integrable cases
in the class (\ref{kdv5par}). Alternatively, in \cite{UGandWH1997a} 
we investigated the conditions on the parameters $\alpha, \beta, \gamma $
such that (\ref{kdv5par}) admits an infinite sequence (perhaps with gaps) 
of polynomial conservation laws. 
The conditions in \cite{UGandWH1997a} are exactly the same as 
the ones above. 

\subsection{Hirota and Satsuma System}

Hirota and Satsuma \cite{RHandJS1981} proposed a coupled system 
of KdV equations,
\begin{eqnarray}\label{hirsatpar}
u_t &=&  6 \alpha u u_x + 6 v v_x + \alpha u_{3x}, \nonumber \\
v_t &=&  3 u v_x + v_{3x},
\end{eqnarray}
where $\alpha$ is a nonzero parameter.
System (\ref{hirsatpar}) describes the interaction of two long waves with
different dispersion relations. It is known to be completely integrable 
provided $\alpha= - \frac{1}{2}.$
The scaling properties of (\ref{hirsatpar}) are such that
$ u \sim v \sim \frac{\partial^2{}}{\partial{x^2}}, \;
\frac{\partial{}}{\partial{t}} \sim \frac{\partial^3{}}{\partial{x^3}}$.
System (\ref{hirsatpar}) is of rank $5$. In a search for the symmetry of 
higher ranks, we obtained the symmetry of rank $7$:
\begin{eqnarray*}
G_1 &=& u^2 u_x - \frac{2}{3} u_x v^2 - \frac{4}{3} u v v_x 
+ \frac{2}{3} u_x u_{2x} - \frac{2}{3} v_x v_{2x} + \frac{1}{3} u u_{3x}
- \frac{2}{3} v v_{3x} + \frac{1}{30} u_{5x}, \\
G_2 &=& -\frac{1}{3} u^2 v_x - \frac{2}{3} v^2 v_x - \frac{1}{3} u_{2x} v_x
- \frac{2}{3} u_x v_{2x} - \frac{2}{3} u v_{3x} - \frac{2}{15} v_{5x},
\end{eqnarray*}
provided that $\alpha = -\frac{1}{2},$ which is the condition for 
complete integrability of (\ref{hirsatpar}). 

\subsection{Volterra Chain}

Consider the following integrable discretization of the KdV equation:
\begin{equation} \label{volterra}
{\dot{u}}_n = u_n \, (u_{n+1}-u_{n-1}),
\end{equation}
which is also known as the Kac-Van Moerbeke equation.
It arises in the study of Langmuir oscillations in plasmas
and in population dynamics \cite{MAandPC1991,MKandPM1975}.

Note that (\ref{volterra}) is invariant under the dilation symmetry
$(t, u_n) \rightarrow (\lambda^{-1} t, {\lambda} u_n)$.
Hence, $u_n$ corresponds to one derivative with respect to $t$,
i.e.\ $u_n \sim \frac{\rm d}{\rm dt}$.

We computed the symmetries of (\ref{volterra}) with ranks $3$ through
$5$. They are:
\begin{eqnarray*}
G^{(1)} &=& u_n u_{n+1} (u_n + u_{n+1} + u_{n+2})
- u_{n-1} u_n (u_{n-2} + u_{n-1} + u_n), \\
G^{(2)} &=&  u_n^3 u_{n+1} + 2 u_n^2 u_{n+1}^2 + u_n u_{n+1}^3 
+ u_n^2 u_{n+1} u_{n+2} + 2 u_n u_{n+1}^2 u_{n+2} \\
& & + u_n u_{n+1} u_{n+2}^2 + u_n u_{n+1} u_{n+2} u_{n+3}
- u_{n-3} u_{n-2} u_{n-1} u_n - u_{n-2}^2 u_{n-1} u_n \\
& & - 2 u_{n-2} u_{n-1}^2 u_n - u_{n-1}^3 u_n - u_{n-2} u_{n-1} u_n^2
- u_{n-1} u_n^3, \\
G^{(3)} &=& u_n^4 u_{n+1} + u_{n-1} u_n^2 u_{n+1}^2 + 3 u_n^3 u_{n+1}^2
+ 3 u_n^2 u_{n+1}^3 + u_n u_{n+1}^4 + u_n^3 u_{n+1} u_{n+2} \\
& & + 4 u_n^2 u_{n+1}^2 u_{n+2} + 3 u_n u_{n+1}^3 u_{n+2} 
+ u_n^2 u_{n+1} u_{n+2}^2
+ 3 u_n u_{n+1}^2 u_{n+2}^2 \\
& & + u_n u_{n+1} u_{n+2}^3 
+ u_n^2 u_{n+1} u_{n+2} u_{n+3} + 2 u_n u_{n+1}^2 u_{n+2} u_{n+3} \\
& & + u_n u_{n+1} u_{n+2} u_{n+3}^2 + u_n u_{n+1} u_{n+2} u_{n+3} u_{n+4}
- u_{n-4} u_{n-3} u_{n-2} u_{n-1} u_n \\
& & - u_{n-3}^2 u_{n-2} u_{n-1} u_n
- 2 u_{n-3} u_{n-2}^2 u_{n-1} u_n - u_{n-2}^3 u_{n-1} u_n \\
& & - 2 u_{n-3} u_{n-2} u_{n-1}^2 u_n - 3 u_{n-2}^2 u_{n-1}^2 u_n
- 3 u_{n-2} u_{n-1}^3 u_n - u_{n-1}^4 u_n \\
& & - u_{n-3} u_{n-2} u_{n-1} u_n^2
- u_{n-2} u_{n-1}^2 u_n^2 - 4 u_{n-2} u_{n-1}^2 u_n^2 - 3 u_{n-1}^3 u_n^2 \\
& & - u_{n-2} u_{n-1} u_n^3 - 3 u_{n-1}^2 u_n^3 - u_{n-1} u_n^4
- u_{n-1}^2 u_n^2 u_{n+1} .
\end{eqnarray*}
Ignoring a trivial misprint in \cite{AMandASandVS1991}, 
Mikhailov {\it et al.\/} listed the symmetry $G^{(1)}.$

\subsection{Ablowitz-Ladik Discretization of the Nonlinear Schr\"odinger 
Equation}

In \cite{MAandJL1976a,MAandJL1976b}, Ablowitz and Ladik studied some of the
properties of the following integrable discretization of the NLS equation:
\begin{equation} \label{orgabllad}
i \, {\dot{u}}_n =
u_{n+1}- 2 u_n + u_{n-1} \pm  u_n^{*} u_n (u_{n+1} + u_{n-1}),
\end{equation}
where $u_n^{*}$ is the complex conjugate of $u_n.$
We continue with the plus sign; the other case would be treated analogously.
Instead of splitting $u_n$ into its real and imaginary parts, we treat
$u_n$ and $v_n = u_n^{*}$ as independent variables and
augment (\ref{orgabllad}) with its complex conjugate equation.
Absorbing $i$ in the scale on $t,$ we get
\begin{eqnarray} \label{abllad}
{\dot{u}}_n
&=& u_{n+1} - 2 u_n + u_{n-1} + u_n v_n (u_{n+1} + u_{n-1}),
\nonumber \\
{\dot{v}}_n
&=& -( v_{n+1} - 2 v_n + v_{n-1} ) -  u_n v_n (v_{n+1} + v_{n-1}).
\end{eqnarray}
Since $v_n = u_n^{*}, $ we have $w(v_n) = w(u_n).$
Neither of the equations in (\ref{abllad}) is uniform in rank.
To circumvent this problem we introduce an auxiliary parameter
$\alpha$ with weight, and replace (\ref{abllad}) by
\begin{eqnarray} \label{ablladnew}
{\dot{u}}_n
&=& \alpha ( u_{n+1} - 2 u_n + u_{n-1} ) + u_n v_n (u_{n+1} + u_{n-1}),
\nonumber \\
{\dot{v}}_n
&=& - \alpha ( v_{n+1} - 2 v_n + v_{n-1} ) -  u_n v_n (v_{n+1} + v_{n-1}).
\end{eqnarray}
Uniformity in rank requires that
\begin{eqnarray*}
w(u_n) + 1 \!&\!=\!&\! w(\alpha) + w(u_n) = 2 w(u_n) + w(v_n) = 3 w(u_n), \\
w(v_n) + 1 \!&\!=\!&\! w(\alpha) + w(v_n) = 2 w(v_n) + w(u_n) = 3 w(v_n),
\end{eqnarray*}
which yields $ w(u_n) = w(v_n) = \frac{1}{2}, w(\alpha) = 1, $
or, $ u_n^2 \sim v_n^2 \sim \alpha \sim {\rm{d} \over \rm{dt}}.$

Recall that the `uniformity in rank' requirement is essential for the first
two steps of the algorithm. However, after step 2, we may set $\alpha = 1.$
The computations now proceed as in the previous examples.
We searched for symmetries of (\ref{abllad}) of ranks $(2,2)$ through
$(7/2,7/2),$ and found symmetries of ranks $(5/2,5/2)$ and
$(7/2,7/2)$. To save space, we only list the symmetries of rank $(5/2,5/2):$
\begin{eqnarray*}
G_1^{(1)} &=& -u_{n+2} - u_n u_{n+1} v_{n-1} - u_{n+1}^2 v_n
- u_n u_{n+2} v_n - u_n^2 u_{n+1} v_{n-1} v_n \\
& & - u_n u_{n+1}^2 v_n^2
- u_{n+1} u_{n+2} v_{n+1} - u_n u_{n+1} u_{n+2} v_n v_{n+1}, \\
G_2^{(1)} &=& v_{n-2} + u_{n-1} v_{n-2} v_{n-1} + u_n v_{n-1}^2 
+ u_n v_{n-2} v_n + u_{n+1} v_{n-1} v_n \\
& & + u_{n-1} u_n v_{n-2} v_{n-1} v_n
+ u_n^2 v_{n-1}^2 v_n + u_n u_{n+1} v_{n-1} v_n^2, \\
G_1^{(2)} &=& -u_{n-2} - u_{n-2} u_{n-1} v_{n-1} - u_{n-1}^2 v_n
- u_{n-2} u_n v_n - u_{n-2} u_{n-1} u_n v_{n-1} v_n \\
& & - u_{n-1}^2 u_n v_n^2
- u_{n-1} u_n v_{n+1} - u_{n-1} u_n^2 v_n v_{n+1}, \\
G_2^{(2)} &=& u_{n-1} v_n v_{n+1} + u_{n-1} u_n v_n^2 v_{n+1} 
+ u_n v_{n+1}^2 + u_n^2 v_n v_{n+1}^2 \\
& & + v_{n+2} + u_n v_n v_{n+2} 
+ u_{n+1} v_{n+1} v_{n+2} + u_n u_{n+1} v_n v_{n+1} v_{n+2} .
\end{eqnarray*}

\subsection{Generalized Toda lattices}

In \cite{YS1997a}, the integrability of the chain
\begin{equation}\label{reltoda}
{\ddot{y}}_n =
\dot{y}_{n+1} e^{(y_{n+1} - y_n)} - e^{2(y_{n+1} - y_n)}
- \dot{y}_{n-1} e^{(y_{n} - y_{n-1})} + e^{2 (y_{n} - y_{n-1})},
\end{equation}
which is related to the relativistic Toda lattice has been studied.
With the change of variables,
$u_n = {\dot{y}}_n, \;  v_n = \exp{(y_{n+1} - y_n)},$
lattice (\ref{reltoda}) can be written as
\begin{equation} \label{reltodalatt}
{\dot{u}}_n =
v_n (u_{n+1} - v_n) - v_{n-1} (u_{n-1} - v_{n-1}), \quad\;\;
{\dot{v}}_n = v_n (u_{n+1} - u_n).
\end{equation}
Here, $ u_n \sim v_n \sim \frac{\rm{d}}{\rm{dt}}. $
We computed a couple of symmetries for (\ref{reltodalatt}). 
One of them reads:
\begin{eqnarray*}
G_1 &=& u_{n-1}^2 v_{n-1} + u_{n-1} u_n v_{n-1} + u_{n-2} v_{n-2} v_{n-1}
- v_{n-2}^2 v_{n-1} - 2 u_{n-1} v_{n-1}^2 \\
& & - u_n v_{n-1}^2 + v_{n-1}^3
- u_n u_{n+1} v_n - u_{n+1}^2 v_n + u_n v_n^2 + 2 u_{n+1} v_n^2 \\
& & - v_n^3 - u_{n+2} v_n v_{n+1} + v_n v_{n+1}^2, \\
G_2 &=& u_n^2 v_n - u_{n+1}^2 v_n + u_{n-1} v_{n-1} v_n - v_{n-1}^2 v_n
- u_n v_n^2 + u_{n+1} v_n^2 \\
& & - u_{n+2} v_n v_{n+1} + v_n v_{n+1}^2 .
\end{eqnarray*}
In \cite{YS1997b}, Suris investigated the integrability of
\begin{equation}\label{backtoda}
{\ddot{y}}_n =
\dot{y}_n \left[\exp{(y_{n+1} - y_n)} - \exp{(y_{n} - y_{n-1})} \right],
\end{equation}
which is closely related to the classical Toda lattice (\ref{orgtoda}).
The same change of variables as for (\ref{reltoda}) allows one to
rewrite (\ref{backtoda}) as
\begin{equation} \label{backtodalatt}
{\dot{u}}_n = u_n (v_n - v_{n-1}), \quad\quad
{\dot{v}}_n = v_n (u_{n+1} - u_n).
\end{equation}
Again, $ u_n \sim v_n \sim \frac{\rm{d}}{\rm{dt}}, $ and
we have computed three symmetries. Two of them are:
\begin{eqnarray*}
G_1^{(1)} &=& u_n^2 v_n + u_n u_{n+1} v_n + u_n v_n^2
- u_{n-1} u_n v_{n-1} - u_n^2 v_{n-1} - u_n v_{n-1}^2, \\
G_2^{(1)} &=& u_{n+1}^2 v_n + u_{n+1} v_n^2 + u_{n+1} v_n v_{n+1}
- u_n^2 v_n - u_n v_{n-1} v_n - u_n v_n^2, \\
G_1^{(2)} &=& u_n^3 v_n + u_n^2 u_{n+1} v_n + u_n u_{n+1}^2 v_n
+ 2 u_n^2 v_n^2 + 2 u_n u_{n+1} v_n^2 + u_n v_n^3 \\
& & + u_n u_{n+1} v_n v_{n+1}
- u_{n-1}^2 u_n v_{n-1} - u_{n-1} u_n^2 v_{n-1} - u_n^3 v_{n-1} \\
& & - u_{n-1} u_n v_{n-2} v_{n-1} - 2 u_{n-1} u_n v_{n-1}^2 
- 2 u_n^2 v_{n-1}^2 - u_n v_{n-1}^3, \\
G_2^{(2)} &=& u_{n+1}^3 v_n - u_n^3 v_n - u_{n-1} u_n v_{n-1} v_n
- 2 u_n^2 v_{n-1} v_n - u_n v_{n-1}^2 v_n - 2 u_n^2 v_n^2 \\
& & + 2 u_{n+1}^2 v_n^2 - u_n v_{n-1} v_n^2 - u_n v_n^3 + u_{n+1} v_n^3
+ 2 u_{n+1}^2 v_n v_{n+1} \\
& & + u_{n+1} u_{n+2} v_n v_{n+1}
+ u_{n+1} v_n^2 v_{n+1} + u_{n+1} v_n v_{n+1}^2 .
\end{eqnarray*}

\section{The Integrability Package}

We now briefly describe the use of our Integrability Package, 
which has (among other things) the code for the computation of symmetries 
based on the algorithms in Sections 2 and 3. 
The Integrability Package is written in {\it Mathematica} \cite{SW1996}
syntax.
Users are assumed to have access to {\it Mathematica} 3.0.
All the necessary files are available in {\it MathSource} \cite{UGandWH1997c} 
including on-line help, documentation, and built-in examples.
The corresponding files should be put in the appropriate places on your
platform. Detailed instructions are given in the documentation. 
\vskip 6pt
\noindent
After launching {\it Mathematica\/}, type
\begin{verbatim}
In[1]:= <<Integrability`
\end{verbatim}
to read in the code. Doing so, you will get the following statement:
\begin{verbatim}
Loading init.m for Integrability from AddOns.
\end{verbatim}
For the purpose of symmetry computations, the functions
{\bf PDESymmetries} and {\bf DDESymmetries} are available.

Working with (\ref{NLSsys}) as an example, the first two lines 
define the system $(r = q^{*})$, whereas the third line will produce 
the three symmetries listed in (\ref{nlssymmetries}): 
\begin{verbatim}
In[2]:= pde1:= D[q[x,t],t] - D[q[x,t],{x,2}] + 
               2*q[x,t]^2*r[x,t] == 0;

In[3]:= pde2:= D[r[x,t],t] + D[r[x,t],{x,2}] - 
               2*r[x,t]^2*q[x,t] == 0;

In[4]:= PDESymmetries[{pde1,pde2},{q,r},{x,t},{4,6},       
            WeightRules->{Weight[q]->Weight[r]}]
\end{verbatim}
Help about the functions and their options can be obtained by typing
\begin{verbatim}
In[5]:= ??DDESymmetries
DDESymmetries[eqn, u, {n, t}, R, opts] finds the symmetry with rank
   R of a differential-difference equation for the function u.
   DDESymmetries[{eqn1, eqn2, ...}, {u1, u2, ...}, {n, t}, R, opts]
   finds the symmetry of a system of differential-difference 
   equations, where R is the rank of the first equation in the
   desired symmetry. DDESymmetries[{eqn1, eqn2, ...},
   {u1, u2, ...}, {n, t}, {Rmax}, opts] finds the symmetries with
   rank 0 through Rmax. DDESymmetries[{eqn1, eqn2, ...},
   {u1, u2, ...}, {n, t}, {Rmin, Rmax}, opts] finds the symmetries
   with rank Rmin through Rmax. n is understood as the discrete
   space variable and t as the time variable.

Attributes[DDESymmetries] = {Protected, ReadProtected}
 
Options[DDESymmetries] = 
  {WeightedParameters -> {}, WeightRules -> {}, 
   MaxExplicitDependency -> 0, UndeterminedCoefficients -> C}
\end{verbatim}
and
\begin{verbatim}
In[6]:= ??WeightedParameters
WeightedParameters is an option that determines the parameters with
   weight. If WeightedParameters -> {p1, p2, ...},
   then p1, p2, .... are considered as constant parameters with
   weight. The default is WeightedParameters -> {}.

Attributes[WeightedParameters] = {Protected}
\end{verbatim}
The option {\bf WeightedParameters} is useful when working with systems that
lack uniformity in rank. In such cases, the code tries to resolve the 
problem of lack of uniformity, and prints appropriate messages.
If the code can not automatically resolve the problem it suggests the use 
of the {\bf WeightedParameters} option. 
Therefore, one should not use the option {\bf WeightedParameters} 
unless it is explicitly suggested. 
For further descriptions of the functions and their
options we refer to the documentation in \cite{UGandWH1997c}.

\section{Other Software Packages}

Higher-order symmetries can be computed with prolongation methods and 
numerous software packages are available that can aid in the tedious 
computations inherent to such methods.
An extensive review of software for Lie symmetry computations, 
including generalized symmetries, can be found in \cite{WH1996,WH1997}. 

With prolongation methods one generates and subsequently reduces and solves
a determining system of linear homogeneous partial differential equations 
for the unknown higher-order symmetry. 
In many cases, due to the length and complexity of that system, 
the general solution is out of reach and one resorts to making 
a polynomial {\it ansatz} for the symmetry. 

Although restricted to polynomial higher-order symmetries, we believe 
that the method presented in this paper is much more straightforward.
Furthermore, it does not require the application of prolongation 
methods or Lie algebraic techniques. 

To avoid retreading the surveys \cite{WH1996,WH1997}, here, we restrict our
discussion to symbolic packages that allow one to compute generalized 
symmetries of PDEs, as they were defined in Section 2.1. 
We are not aware of software for DDEs to calculate the type of symmetries 
defined in Section 3.1.

Based on the alternative strategy discussed in Remark (ii) in Section 2, 
Ito's programs in REDUCE \cite{MI1986,MI1994} compute polynomial 
higher-order symmetries for systems of evolution equations that are 
uniform in rank (no weighted parameters can be introduced).
Ito programs can not be used to computed symmetries that explicitly 
depend on $x$ and $t.$ 

In \cite{BFandSIandWW1997}, Fuchssteiner {\it et al.\/} present an 
algorithm to compute higher-order symmetries of evolution 
equations. Their algorithm is based on Lie algebraic techniques 
and uses commutator algebra on the Lie algebra of vector fields. 
Their approach is different from the usual prolongation method in that
no determining equations are solved. Instead, all necessary generators 
of the finitely generated Virasoro algebra are computed from one 
given element by direct Lie algebra methods.
Their code is available in MuPAD. 

The REDUCE program {\bf FS} for ``formal symmetries'' was written by 
Gerdt and Zharkov \cite{VGandAZ1990}.
The code {\bf FS} can be applied to polynomial nonlinear PDEs of evolution 
type, which are linear with respect to the highest-order spatial 
derivatives and with non-degenerated, diagonal coefficient matrix
for the highest derivatives. The algorithm in {\bf FS} requires that the 
evolution equation are of order two or higher in the spatial variable.
However, this approach does not require that the evolution
equations are uniform in rank. 
With {\bf FS} one cannot compute symmetries that depend explicitly 
on $x$ and $t.$ 
%

The PC package {\bf DELiA}, written in Turbo PASCAL by Bocharov 
\cite{AB1991} and co-workers, is a commercial computer algebra system 
for investigating differential equations using Lie's approach.
The program deals with higher-order symmetries, conservation laws, 
integrability and equivalence problems. 
It has a special routine for systems of evolution equations.
The program requires the presence of second or higher-order spatial 
derivative terms in all equations.

Finally, Sanders and Wang \cite{JSandJPW1997} have Maple and FORM software 
that aids in the computation of recursion operators.

\section{Conclusions}

We have implemented direct algorithms (in {\it Mathematica}) that 
allow the user to compute polynomial higher-order symmetries of 
polynomial systems of evolution and lattice equations. 

These algorithms are based on the dilation invariance of the given equations. 
Only minor modifications of our strategy lead to direct algorithms for
conserved densities for systems of nonlinear PDEs 
\cite{UGandWH1997a} and nonlinear DDEs \cite{UGandWH1997b,UGandWHandGE1997}.

For systems that arise from a variational principle, conservation laws 
follow from higher-order symmetries (Noether's theorem) and vice versa. 
Currently, in our algorithms we are not exploiting such connections.

In the future we will investigate generalizations of our methods to 
PDEs and DDEs in multiple space dimensions. We will also study the 
potential use of Lie-point symmetries other than dilation symmetries.
Moreover, most recursion operators, which connect generalized symmetries, 
are also dilation invariant. We are extending our algorithms to the 
symbolic computation of recursion operators.

\section*{Acknowledgements}

We acknowledge helpful discussions with Profs. Clara Nucci, Jan Sanders,  
and Pavel Winternitz.


\begin{thebibliography}{99}

\bibitem{MAandPC1991}
M. J. Ablowitz and P. A. Clarkson,
{\em Solitons, Nonlinear Evolution Equations and Inverse Scattering},
Cambridge University Press, Cambridge, 1991.

\bibitem{MAandJL1976a}
M. J. Ablowitz and J. F. Ladik,
Nonlinear differential-difference equations and Fourier analysis,
J. Math. Phys., 17 (1976) 1011--1018.

\bibitem{MAandJL1976b}
M. J. Ablowitz and J. F. Ladik,
A nonlinear difference scheme and inverse scattering,
Stud. Appl. Math., 55 (1976) 213--229.


\bibitem{AB1991}
A. V. Bocharov,
{\em DELiA: a system for exact analysis of Differential Equations using
S. Lie Approach}, DELiA 1.5.1 User Guide, 
Beaver Soft Programming Team, New York (1991).

\bibitem{RDandJG1977}
R. K. Dodd and J. D. Gibbon, 
The prolongation structure of a higher order Korteweg-de Vries equation,
Proc. R. Soc. Lond. A,  358 (1977) 287--296.

\bibitem{NEandWHS1992}
N. Euler and W. H. Steeb,
{\em Continuous Symmetries, Lie Algebras and Differential Equations},
Wissenschaftsverlag, Mannheim, 1992.

\bibitem{AF1980}
A. S. Fokas,
A symmetry approach to exactly solvable evolution equations,
J. Math. Phys., 21 (1980) 1318--1325.

\bibitem{AF1987}
A. S. Fokas,
Symmetries and integrability,
Stud. Appl. Math., 77 (1987) 253--299.

\bibitem{AFandJG1980}
A. Fordy and J. Gibbons, 
Some remarkable nonlinear transformations,
Phys. Lett. A, 75 (1980) 325.

\bibitem{BFandSIandWW1997}
B. Fuchssteiner, S. Ivanov and W. Wiwianka,
Algorithmic determination of infinite-dimensional symmetry groups 
for integrable systems in 1+1 dimensions,
Mathl. Comput. Modelling, 25 (1997) 91--100.

\bibitem{BFandWOandWW1987}
B. Fuchssteiner, W. Oevel and W. Wiwianka,
Computer-algebra methods for investigation of hereditary operators of 
higher order soliton equations,
Comput. Phys. Commun., 44 (1987) 47--55.

\bibitem{VGandAZ1990}
V. P. Gerdt and A. Y. Zharkov,
Computer generation of necessary integrability conditions for
polynomial-nonlinear evolution systems,
in {\it Proc. ISSAC '90\/}, S. Watenabe and M. Nagata eds., 
Academic Press, New York, 1990, pp. 250--254.

\bibitem{UGandWH1997a} 
\"{U}. G\"{o}kta\c{s} and W. Hereman,
{\em Symbolic computation of conserved densities for systems of nonlinear
evolution equations},
J. Symbolic Computation, 24 (1997) 591--621.

\bibitem{UGandWH1997b} 
\"{U}. G\"{o}kta\c{s} and W. Hereman,
Computation of conservation laws for nonlinear lattices, 
Physica D (1998) to appear.

\bibitem{UGandWH1997c}
\"U. G\"okta\c{s} and W. Hereman,
The software package and the related files are available at 
http://www.mathsource.com/cgi-bin/msitem?0208-932.
{\it MathSource\/} is a vast electronic library of {\it Mathematica\/} 
material.

\bibitem{UGandWHandGE1997}
\"{U}. G\"{o}kta\c{s}, W. Hereman and G. Erdmann,
Computation of conserved densities for systems of nonlinear 
differential-difference equations, 
Phys. Lett. A, 236 (1997) 30-38.

\bibitem{MH1974}
M. H\'{e}non,
Integrals of the Toda lattice,
Phys. Rev. B, 9 (1974) 1921--1923.

\bibitem{WH1996}
W. Hereman,
Symbolic software for Lie symmetry analysis,
in: CRC Handbook of Lie Group Analysis of Differential Equations, 
Volume 3: New Trends in Theoretical Developments and Computational Methods, 
Chapter 13, Ed.: N.H. Ibragimov, 
CRC Press, Boca Raton, Florida (1996) 367-413. 

\bibitem{WH1997}
W. Hereman,
Review of symbolic software for Lie symmetry analysis,
Mathl. Comput. Modelling, 25 (1997) 115-132.

\bibitem{RHandMI1983}
R. Hirota and M. Ito,
Resonance of solitons in one dimension,
J. Phys. Soc. Jpn., 52 (1983) 744--748.

\bibitem{RHandJS1981}
R. Hirota and J. Satsuma,
Soliton solutions of a coupled Korteweg-de Vries equation,
Phys. Lett. A, 85 (1981) 407--408.

\bibitem{MI1986}
M. Ito,
A REDUCE program for finding symmetries of nonlinear evolution 
equations with uniform rank,
Comput. Phys. Commun., 42 (1986) 351--357. 

\bibitem{MI1994}
M. Ito, 
SYMCD - a REDUCE package for finding symmetries and conserved densities
of systems of nonlinear evolution equations,
Comput. Phys. Commun., 79 (1994) 547--554.

\bibitem{MKandPM1975}
M. Kac and P. van Moerbeke,
On an explicitly soluble system of nonlinear differential equations
related to certain Toda lattices,
Adv. Math., 16 (1975) 160--169.

\bibitem{DK1980}
D. J. Kaup, 
On the inverse scattering problem for cubic eigenvalue problems of the
class $\Psi_{xxx}+ 6 Q \Psi_{x} + 6 R \Psi = \lambda \Psi\/$,
Stud. Appl. Math., 62 (1980) 189--216.

\bibitem{BKandGW1981}
B. A. Kupershmidt and G. Wilson, 
Modifying Lax equations and the second Hamiltonian structure,
Invent. Math., 62 (1981) 403--436.

\bibitem{PL1968}
P. D. Lax,
Integrals of nonlinear equations of evolution and solitary waves,
Commun. Pure Appl. Math., 21 (1968) 467--490.

\bibitem{AMandASandVS1991}
A. V. Mikhailov, A. B. Shabat and V. V. Sokolov,
The symmetry approach to classification of integrable equations,
in {\em What Is Integrability?}, V. E. Zakharov ed., 
Springer-Verlag, Berlin Heidelberg, 1991, pp. 115--184.

\bibitem{PO1977}
P. J. Olver,
Evolution equations possessing infinitely many symmetries,
J. Math. Phys., 18 (1977) 1212--1215.

\bibitem{PO1993}
P. J. Olver, 
{\em Applications of Lie Groups to Differential Equations,} 2nd Edition,
Springer Verlag, New York, 1993.

\bibitem{ARandBGandKT1992}
A. Ramani, B. Grammaticos and K. M. Tamizhmani,
An integrability test for differential-difference systems,
J. Phys. A: Math. Gen., 25 (1992) L883--L886.

\bibitem{VRandGK1994}
V. Rosenhaus, G. H. Katzin,
On symmetries, conservation laws, and variational problems for partial
differential equations,
J. Math. Phys., 35 (1994) 1998--2012.

\bibitem{JSandJPW1997}
J. Sanders and J. P. Wang, 
On hereditary recursion operators,
Report WS-472, Department of Mathematics and Computer Sciences, 
Free University, Amsterdam, The Netherlands,
Physica D (1997) submitted.

\bibitem{JSandDK1977}
J. Satsuma and D. J. Kaup, 
A B\"{a}cklund transformation for a higher order Korteweg-De Vries equation,
J. Phys. Soc. Jpn., 43 (1977) 692--697.

\bibitem{KSandTK1974}
K. Sawada and T. Kotera,
A method for finding N-Soliton solutions of the K.d.V. equation and
K.d.V.-like equation,
Prog. Theor. Phys., 51 (1974) 1355--1367.

\bibitem{YS1997a}
Y. B. Suris,
New integrable systems related to the relativistic Toda lattice,
J. Phys. A: Math. Gen., 30 (1997) 1745--1761.

\bibitem{YS1997b}
Y. B. Suris,
On some integrable systems related to the Toda lattice,
J. Phys. A: Math. Gen., 30 (1997) 2235--2249.

\bibitem{MT1981}
M. Toda,
{\em Theory of nonlinear lattices},
Springer Verlag, Berlin, 1981.


\bibitem{SW1996}
S. Wolfram,
{\em The Mathematica book}, 3rd Edition,
Wolfram Media, Urbana-Champaign, Illinois \&
Cambridge University Press, London, 1996.

\end{thebibliography}
\end{document}